# HIGH LUMINOSITY ISSUES FOR DAΦNE UPGRADE

F. Ruggiero, CERN, Geneva, Switzerland

M. Zobov, INFN-LNF, Frascati, Italy

*Abstract*

We give an overview of presentations and discussions during the Accelerator Working Group Session dedicated to High Luminosity Issues for a future upgrade of the Frascati $e^+e^-$ Φ – Factory DAΦNE at the Workshop "$e^+e^-$ in the 1-2 GeV range: Physics and Accelerator Prospects" held at Alghero (Italy) on 10-13 September 2003.

## 1. INTRODUCTION

The first Session of the Accelerator Working Group was dedicated to High Luminosity Issues of the DAΦNE upgrade. The Session was entitled "Beam-Beam Interaction", and included presentations by C. Biscari, M. Zobov, H. Ikeda, P. Raimondi, A. Gallo, and A. Temnykh. The main goal of the discussion was to focus on some new ideas that may lead to a substantial luminosity increase and also to share the experience with other Particle Factories, rather than to review beam-beam effects in detail.

## 2. PARTICLE FACTORIES PLANS

To initiate the discussion, C. Biscari presented an overview talk "Upgrade of Particle Factories" [1] describing several ideas and plans for the luminosity increase. She clearly showed that a new era for Particle Factories has started:

- Both B-Factories, KEKB and PEP-II, have successfully overcome their design luminosities and, at present, have upgrade plans with an ultimate luminosity goal of $10^{36}$ cm$^{-2}$ s$^{-1}$ which has to be reached in steps (see Fig. 1).

- The CESR collider will explore physics at lower energies, thus becoming a "τ-charm Factory".

- Upgrade of the Beijing BEPC into BEPC-II will bring the Chinese collider to the particle Factory level in the energy range of 2-5.6 GeV.

- The "Light Quark Factory" VEPP2000 is under construction in Novosibirsk and it is scheduled to start commissioning at the end of next year.

- The upgrade of the Φ-Factory DAΦNE is the main subject of the present Workshop.

The Particle Factories plans with their time schedules are summarized in Figure 1. More details can be found in [1, 2] and references therein.

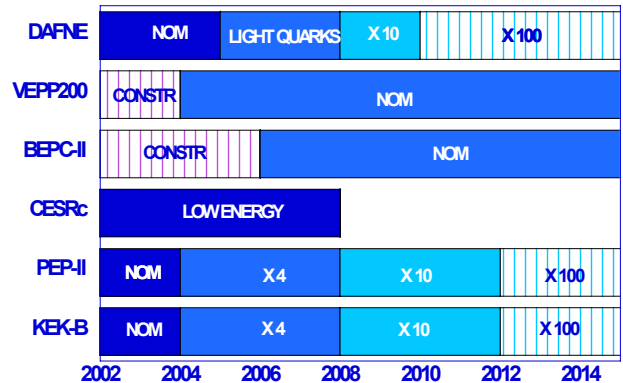

Fig. 1 Timetable of foreseen Particle Factories upgrade (C. Biscari [1]).

In addition to conventional ways to increase the machine luminosity, such as shorter bunches, smaller interaction point beta functions, more bunches, higher current, higher RF frequencies etc., the following ideas are under consideration and have yet to be tested:

- round beams (VEPP-2000);

- crab-crossing collisions (KEKB);

- four beams compensation scheme (KEKB);

- lattice with a negative momentum compaction (DAΦNE, KEKB, BEPC-II).

The DAΦNE Team has come up with three new ideas described in the talks of M. Zobov, P. Raimondi and A. Gallo during this Session.

## 3. NEGATIVE MOMENTUM COMPACTION

The DAΦNE lattice is flexible enough to provide collider operation with a negative momentum compaction factor $\alpha_C$. This can be considered as an intermediate step in the DAΦNE luminosity upgrade without substantial hardware changes (M. Zobov [3]). There can be several advantages for beam dynamics and luminosity performance in a collider with negative momentum compaction:

- Bunches are shorter, with a more regular shape. Shorter bunches are preferable for both peak luminosity increase and beam lifetime improvement, since the transverse beta functions at the IP can be reduced without limitations due to the hour-glass effect and, in addition, the Piwinski angle is lower in collisions with a crossing angle, as in the case of DAΦNE and KEKB.

- Longitudinal beam-beam effects are less dangerous. In particular, the coherent and incoherent instabilities due to the longitudinal kick during beam-beam collisions do not take place with a negative momentum compaction. Moreover, the beam-beam synchrobetatron resonances are less harmful in this case [4].

- The microwave instability threshold is higher [5]. (However, as the experience shows, this depends much on the actual machine coupling impedance [6,7]).

- Since the head-tail instability with a negative momentum compaction takes place with positive chromaticity, the requirements on the sextupole strength can be relaxed. Indeed, in the SUPER-ACO it has been possible to store 100 mA in a single bunch without sextupoles [7].

Numerical simulations of the bunch lengthening in DAΦNE with negative momentum compaction were based on the calculated wake potential, which has already been successfully applied for bunch lengthening and microwave threshold predictions [8]. It has been shown that for $\alpha_C = -0.024$ it is possible to keep the bunch length below 1.5 cm up to a bunch current of 30 mA. This will allow reducing the vertical beta function at the IP down to 1.5 cm.

Subsequent numerical simulations of beam-beam effects with LIFETRAC [9], taking into account the calculated bunch length, indicate that by shifting the working point close to an integer and switching to a lattice with negative momentum compaction, it is possible to push DAΦNE luminosity up to the $10^{33}$ cm$^{-2}$ s$^{-1}$ level.

The idea of a negative momentum compaction seems to be quite obvious and transparent, since it has been proposed independently for KEKB [10] and it is also under consideration for BEPC-II (C. Zhang [11]). Moreover, the KEKB team managed to implement an experimental lattice with negative compaction and measured the bunch length just a few days before the Summer 2003 collider shut down (H. Ikeda [12]).

The bunch length has been measured by three different methods, namely detection of two different frequencies of the bunch spectrum by a button electrode, beam-induced fields in an RF wave-guide, and streak camera. The results consistently show a bunch length reduction in both HER and LER rings with a negative momentum compaction lattice.

An example of the measurement with the RMS bunch length button monitor is shown in Fig. 2. It is possible to observe a typical bunch length behaviour with the negative momentum compaction: below the microwave instability threshold, the bunch gets shorter due to the potential well distortion and the bunch length starts growing above the threshold.

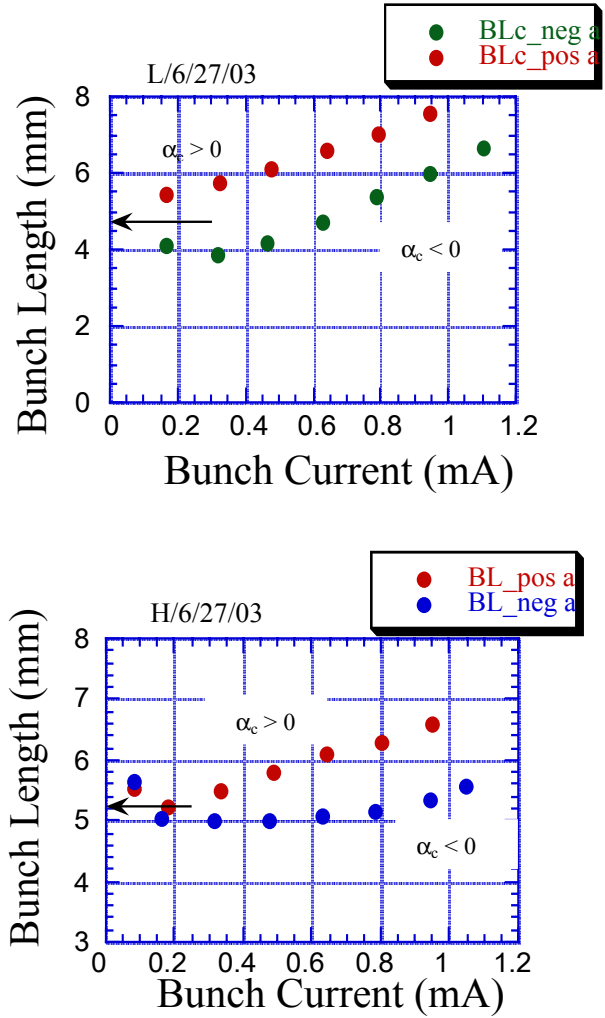

Fig. 2 Bunch lengthening in KEKB LER (upper plot) and HER (lower plot) as measured with rms bunch length monitor (H. Ikeda [12]).

## 4. COLLISIONS UNDER LARGE CROSSING ANGLE

P. Raimondi proposed to use high energy beams colliding at a large crossing angle to get low energy in the center of mass: $E_{cm} = 2 E \cos(\phi)$ [13]. In this case, several advantages of high energy colliders can be exploited for the luminosity increase at the low center of mass energy. The most relevant ones are:

- Tune shifts scale with 1/Energy (E), leading to a fundamental quadratic increase of the luminosity versus energy provided that the number of particles per bunch scales linearly with E.

- Radiation damping times decrease with $1/E^3$ leading to higher limits for tune shifts.
- Touschek effect decreases with $1/E^3$.
- Natural bunch length is shorter.
- Beam is stiffer: single and multi-bunch instabilities growth times increase with E.

Possible significant advantages can also come from:

- A simple and more flexible IR design. Due to the large crossing angle, the final focus quadrupoles can be moved closer to the IP, thus providing smaller betas at the IP with lower chromaticity;
- Kaons will be boosted, so it may be possible to have the detector decoupled from the IR.
- Reversing the direction of one of the beams, it may be possible to increase the center of mass energy thus allowing the high energy solution as well.

There are two possible options in the scheme with large crossing angle: with and without crab-crossing.

- In the "no crab-crossing case" the luminosity is reduced by the factor $\sigma_x/(\sigma_z \tan(\phi))$. However, the tune shifts and the interaction length are also reduced, leading to smaller design emittances and vertical beta function at the IP. For this option the shorter bunch length and stronger damping would be particularly preferable in order to reduce the Piwinski angle and to cope with beam-beam synchro-betatron resonances [14]. In addition, with shorter bunches the luminosity reduction factor becomes smaller.
- The "crab-crossing" is still possible at the relatively low energies under consideration, but usage of superconducting cavities is mandatory. The luminosity does not change with the crossing angle, however, the interaction length at the IP is longer, leading to an increase of the minimum vertical beta with a corresponding luminosity loss.

According to the luminosity expectation at the Φ-resonance energy, a luminosity within a factor of 4 around the $10^{34}$ cm$^{-2}$ s$^{-1}$ range could be obtained with this new scheme.

## 5. STRONG RF FOCUSING

The innovative idea of a strong RF focusing in a collider [15] was presented at the Session by A. Gallo [16]. As it is well known, the luminosity of a circular collider can be increased by decreasing the bunch length and the beta functions. A natural way to decrease the bunch length is to decrease the momentum compaction factor and/or to increase the RF voltage. However, in such a way short bunches cannot be obtained with high beam currents, since wake fields prevent this through the potential well distortion and the microwave instability.

The proposed strong RF focusing (with high RF voltage and high momentum compaction factor) can be used to focus longitudinally the bunch at the IP with its progressive lengthening towards the RF cavity. This allows placing the major sources of impedance near the RF cavity, where the bunch is longest, thus minimizing the effect of the wake fields.

As it was shown the bunch length excursion depends strongly on the longitudinal phase advance. When the phase advance tends to 180 degrees, the ratio of the bunch length at the RF cavity position and at the IP goes to infinity. An important feature of the strong RF focusing is that the equilibrium energy spread also depends on the longitudinal phase advance. Figure 3 shows a good agreement between the bunch length and the energy spread obtained analytically and by numerical simulations as a function of the phase advance. Finally, a hypothetical set of parameters for a Φ-factory aimed at a luminosity of $10^{34}$ cm$^{-2}$ s$^{-1}$ was analysed, with the conclusion that it is possible to obtain a bunch length of 2 mm at the IP.

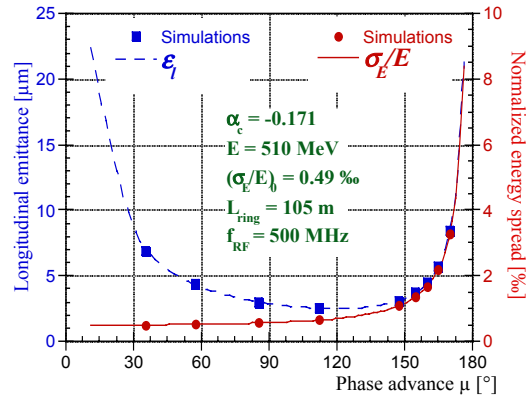

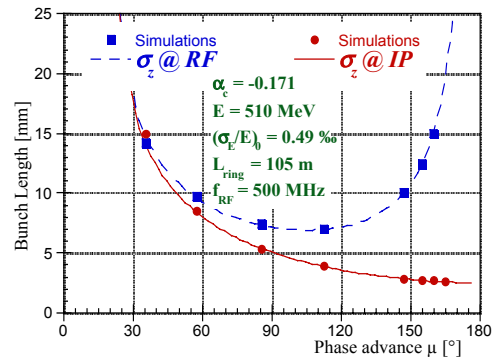

Fig. 3 Longitudinal emittance (upper figure, left scale), relative energy spread (upper figure, right scale), and bunch length (lower figure) as a function of the longitudinal phase advance: lines – analytical results; dots – numerical simulations (A. Gallo [16]).

## 6. BEAM-BEAM EXPERIENCE AT VEPP-4

The right choice of working point is crucial for a successful collider performance.

A. Temnykh presented the results of his experimental study of the luminosity dependence on the working point position for VEPP-4 [17]. A 2D tune scan technique has been used to explore the tune plane measuring the vertical beam size from luminosity values and positron beam loss rates. A variety of resonances excited by machine nonlinearity and beam-beam interaction affecting the luminosity and the beam lifetime was observed and identified. The tune scan has suggested an optimal working point close to half-integer tunes. This conclusion is also in accordance with the experience of PEP-II, KEKB and CESR.

Another subject presented by A. Temnykh is the influence of the machine nonlinearities on the beam-beam interaction [17]. By applying an analytical model including beam-beam effects and a cubic machine nonlinearity (octupoles), he has shown that the cubic nonlinearity may dramatically change the beam-beam interaction dynamics. In this sense not only the strength of the cubic nonlinearity is important, but also its sign. This is also confirmed by the DAΦNE experience [18]. Experimentally it was found that for VEPP-4 the cubic nonlinearity must be kept close to zero for optimum collider performance and that a negative nonlinearity may be a little bit less harmful than a positive one.

## 7. CONCLUSIONS

In conclusion, several new ideas to substantially increase machine luminosity can and will be tested in the near future:

- Crab cavities (KEK-B);
- Collisions with round beams (VEPP2000);
- Negative momentum compaction and strong damping (KEK-B, DAΦNE);
- Strong RF focusing (CESR?).

The approach of the DAΦNE machine team is sound, since a luminosity of $10^{34}$ cm$^{-2}$ s$^{-1}$ is already a challenging target. Reaching a luminosity of $10^{35}$ cm$^{-2}$ s$^{-1}$ needs many combined new ideas/technologies and is thus associated with higher risks and a longer time scale.